\documentclass[11pt]{article}
\usepackage[dvips]{graphicx}
\usepackage{amsmath,amssymb,epsfig,bm,slashbox}

\begin{document}

\date{\empty}
\title{\textbf{Do intergalactic magnetic fields imply}\\ \textbf{an open universe?}}

\author{John D. Barrow${}^{1}$, Christos G. Tsagas${}^{2}$ and Kei Yamamoto${}^{1}$\\ {\small ${}^1$DAMTP, Centre for Mathematical Sciences, University of Cambridge}\\ {\small Wilberforce Road, Cambridge CB3 0WA, UK}\\ {\small ${}^2$Section of Astrophysics, Astronomy and Mechanics, Department of Physics}\\ {\small Aristotle University of Thessaloniki, Thessaloniki 54124, Greece}}

\maketitle

\begin{abstract}
The detection of magnetic fields at high redshifts, and in empty intergalactic space, support the idea that cosmic magnetism has a primordial origin. Assuming that Maxwellian electromagnetism and general relativity hold, and without introducing any `new' physics, we show how the observed magnetic fields can easily survive cosmological evolution from the inflationary era in a marginally open Friedmann universe but fail to do so, by a very wide margin, in a flat or a marginally closed universe. Magnetic fields evolve very differently in open and closed Friedmann models. The existence of significant magnetic fields in the universe today, that require primordial seeding, may therefore provide strong evidence that the universe is marginally open rather than marginally closed.
\end{abstract}

The galactic dynamo could in principle have amplified magnetic seeds as weak as $10^{-20}$~G to the $O(\mu$G$)$ fields seen in galaxies today, if these seeds are coherent on lengths comparable to that of the smallest proto-galactic eddy, which is about 10~Kpc in comoving scale~\cite{KA}. However, the efficiency of the dynamo mechanism is under scrutiny. Also, galactic dynamos find it difficult to explain the protogalactic magnetic fields detected at high redshifts with strengths between $10^{-7}$ and $10^{-6}$~G, which are very close to those observed in fully formed nearby galaxies. Yet, the greatest puzzle comes from recent surveys reporting magnetic fields in empty intergalactic space, where a dynamo presumably cannot operate, with magnitudes around $10^{-15}$~G~\cite{TGFBGC}. These surveys account for the magnetic effects on GeV $\gamma$-rays, produced by the interaction between TeV-energy photons from distant Active Galactic Nuclei (AGN) and low-frequency background photons. The overall result of a magnetic presence in intergalactic space is an extended halo around the $\gamma$-ray image of these AGNs. In addition, the $B$-field deflects the $\gamma$-rays and so reduces their observed flux. So far, the data seem consistent with intergalactic magnetic strengths between $10^{-17}$ and $10^{-14}$~G. What kind of mechanism can produce $B$-fields of these strengths?

The fate of cosmological magnetic fields depends crucially on whether the universe has negatively or positively curved spatial sections. Cosmological magnetic fields evolving on flat or closed Friedmann backgrounds become negligibly small by the present, but even marginally open Friedmann models can sustain astrophysically relevant $B$-fields. The WMAP results support a nearly flat universe (with $|\Omega_{0}-1|\lesssim10^{-2}$) but stop short from favouring a slightly open or a slightly closed one~\cite{Setal}. The dependence of the magnetic survival on the spatial curvature is so strong, that it may be used to determine whether the visible universe has open or closed space sections. We suggest that, if the observed magnetic fields are relics from the inflationary era, then we need to have $\Omega_{0}<1$, an open universe.

This sensitivity to whether the geometry is open or closed in the $\Omega\rightarrow1$ limit is also familiar from bounds on the shear anisotropy of the universe, under the assumption that the anisotropy is contributed by homogeneous, Bianchi-type gravitational-wave modes. Those limits reveal that the bounds on shear and vorticity contributed by long-wavelength homogeneous gravitational waves in closed universes are much stronger than those bounds for flat or open universes~\cite{CH}. The finite-wavelength transverse tensor modes are more strongly constrained and a similar effect can be seen when compact topologies are imposed on the space sections of the flat and open universes and periodic boundary conditions need to be satisfied~\cite{BK}. When magnetic fields are present the spatial curvature also plays an important role through a magnetic vector coupling to the spacetime geometry~\cite{E}.

Inflationary magnetic fields are thought to be too weak to seed the galactic dynamo, unless conventional electromagnetism, or standard cosmology, is abandoned, because of the belief that $B$-fields on Friedmann backgrounds always decay as $B\propto a^{-2}$, where $a$ is the cosmological scale factor. To demonstrate this, let us rescale the magnetic vector as $\mathcal{B}_{a}=a^{2}B_{a}$ and use conformal, rather than proper, time. Then, to linear order, we have~\cite{GR}
\begin{equation}
\mathcal{B}_{(n)}^{\prime\prime}+ n^{2}\mathcal{B}_{(n)}= 0\,,  \label{fFRW}
\end{equation}
where $B_{(n)}$ represents the $n$-th magnetic mode and primes indicate conformal time derivatives. Note that the above holds in cosmological environments of poor electrical conductivity, like those of a typical inflationary phase, or on scales beyond the Hubble horizon after inflation.\footnote{Once inflation ends, the conductivity of the universe increases rapidly and the emerging currents quickly disperse the electric fields and freeze their magnetic counterparts into the primordial plasma. Nevertheless, causality ensures that there are no currents outside the Hubble radius, which implies that expression (\ref{fFRW}) still holds there. Note that in highly conducting environments, cosmological magnetic fields always decay adiabatically.} Equation (\ref{fFRW}) gives an oscillatory solution for $\mathcal{B}_{(n)}$ with constant amplitude. This implies that $B_{(n)}\propto a^{-2}$ on all scales, irrespective of the type of matter that fills the universe, and a residual magnetic field strength today below $10^{-50}$~G~\cite{GR}. Magnetic seeds like these are astrophysically irrelevant.

However, the Minkowski-like wave equation (\ref{fFRW}) only holds on \textit{flat} Friedmann backgrounds. When the 3-D hypersurfaces have non-Euclidean geometry, expression (\ref{fFRW}) becomes~\cite{TK}
\begin{equation}
\mathcal{B}_{(n)}^{\prime\prime}+ \left(n^{2}+2K\right)\mathcal{B}_{(n)}= 0\,,  \label{cFRW}
\end{equation}
with $K=0,\pm1$ for the 3-curvature index of the unperturbed Friedmann model.\footnote{The Laplacian eigenvalue ($n$) takes continuous values, with $n^{2}\geq0$, in flat and open cosmologies. When the spatial geometry is spherical the eigenvalue is discrete, with $n^{2}\geq3$.} The curvature term in Eq.~(\ref{cFRW}) is a purely general relativistic effect, arising from the vector nature of the gravitating electromagnetic field~\cite{TK}.

One of the implications of the magneto-curvature coupling seen in Eq.~(\ref{cFRW}), is that $B$-fields do not necessarily decay adiabatically on all Friedmann backgrounds, and their evolution depends critically on the spatial curvature of the host. The differences appear on relatively large scales, since on sufficiently small lengths (i.e.~for modes with $n^{2}\gg2$) the three versions of relation (\ref{cFRW}) are essentially identical. This is not surprising, given that the direct 3-curvature effects are expected to fade away as we move on to progressively smaller wavelengths. Another implication of Eq.~(\ref{cFRW}) is the absence of a real change in the magnetic evolution between the flat and the closed Friedmann models. Indeed, when $K=0,+1$, the solution of (\ref{cFRW}) reads
\begin{equation}
\mathcal{B}_{(n)}= \mathcal{C}_{1}\cos\left(\eta\sqrt{2K+n^{2}}\right)+ \mathcal{C}_{2}\sin\left(\eta\sqrt{2K+n^{2}}\right)\,,  \label{Bn1}
\end{equation}
which means that the rescaled $\mathcal{B}$-field oscillates with constant amplitude. Consequently, the actual magnetic field decays as $a^{-2}$ in both cases (recall that $\mathcal{B}=a^{2}B$ by definition). The only difference the positive curvature makes is in the frequency of the magnetic oscillation and this is only noticeable on relatively large scales (i.e.~for small values of $n$).

The situation changes drastically when the Friedmann background has
hyperbolic spatial geometry, in which case $K=-1$ and Eq.~(\ref{cFRW}) recasts into
\begin{equation}
\mathcal{B}_{(n)}^{\prime\prime}+ \left(n^{2}-2\right)\mathcal{B}_{(n)}= 0\,.  \label{oFRW}
\end{equation}
Although short wavelengths still oscillate with a constant amplitude, on sufficiently large scales there is a significant qualitative change. When $0\leq n^{2}<2$, expression (\ref{oFRW}) no longer accepts conventional wave solutions, but ``exponential waves'' of the form
\begin{equation}
\mathcal{B}_{(n)}= \mathcal{C}_{1}\cosh\left(\eta\sqrt{2-n^{2}}\right)+ \mathcal{C}_{2}\sinh\left(\eta\sqrt{2-n^{2}}\right)\,.  \label{cBn}
\end{equation}
In terms of the cosmological scale factor, the above implies the following evolution-law for the actual magnetic field~\cite{TK}
\begin{equation}
B_{(n)}= C_{1}\left({\frac{a}{a_{0}}}\right)^{\sqrt{2-n^{2}}-2}+ C_{2}\left({\frac{a}{a_{0}}}\right)^{-\sqrt{2-n^{2}}-2}\,,  \label{Bn2}
\end{equation}
where $0\leq n^{2}<2$ from now on. This holds for a period of slow-roll inflation, during reheating, and subsequently in the radiation and dust epochs. Throughout this time $B$-fields spanning scales close to and beyond the curvature radius of an open Friedmann model, which corresponds to $n^{2}=1$, are \textit{superadiabatically amplified} by curvature effects alone.

This is possible, despite the conformal invariance of Maxwellian electromagnetism and the conformal flatness of the Friedmann universes. The reason is the nature of the conformal mapping between the Minkowski spacetime and the three Friedmann models, which changes depending on the spatial geometry of the latter. For the flat model the mapping is global, but for the other two is local and breaks down on sufficiently large scales where the curvature starts to dominate~\cite{S}. The global nature of the conformal relation between the flat Friedmann cosmology and the Minkowski space guarantees the rapid adiabatic decay of the actual $B$-field on all scales. The local nature of the conformal mapping between the curved Friedmann models and the Minkowski spacetime, implies that on large enough scales, where curvature dominates, the Minkowski-like evolution for the rescaled $\mathcal{B}$-field no longer holds. There, one has to switch from Eq.~(\ref{fFRW}) to expression (\ref{cFRW}). As a result, the adiabatic magnetic decay is not a priori guaranteed for all Friedmann universes because of spatial curvature effects. These do not seem to make any real difference when the background model is closed, but can drastically change the magnetic evolution in the case of an open universe. In fact, the ability to superadiabatically amplify large-scale magnetic fields appears to be a generic property of universes with hyperbolic spatial geometry, since analogous effects have also been observed in open Bianchi class B models~\cite{Y}.

Following solution (\ref{Bn2}) at the curvature scale, magnetic fields decay as $a^{-1}$, instead of dropping at the adiabatic $a^{-2}$ rate. This can lead to residual $B$-seeds considerably stronger than in the zero curvature case. Evolving solution (\ref{Bn2}) throughout the universe's lifetime, one can show that the current magnetic strength depends on the energy scale of the adopted inflationary scenario and on the present density parameter of the universe~\cite{TK}. Specifically, we find
\begin{equation}
B_{0}\sim 10^{-65+51\sqrt{2-n^{2}}} \left({\frac{M}{10^{14}GeV}}\right)^{2\sqrt{2-n^{2}}} \left[(1-\Omega_{0})(n^{2}-1)\right]^{(2-\sqrt{2-n^{2}})/2} \hspace{5mm} \mathrm{G}\,,  \label{B01}
\end{equation}
for modes coherent on the largest subcurvature lengths (i.e.~with $1<n^{2}<2$).\footnote{Expression (\ref{B01}) applies only to $B$-fields coherent on subcurvature scales. The latter lie inside the Hubble horizon and cross outside during the late de Sitter phase of the inflationary expansion~\cite{TK}. For these magnetic modes the time of horizon crossing and the subsequent number of e-folds are crucial because they determine the strength of the residual field. Supercurvature modes, with $0<n^{2}\leq1$, are also superadiabatically amplified (see Eq.~(\ref{Bn2})). However, the corresponding scales are always outside the Hubble radius of an open Friedmann universe. For these modes, there is no horizon crossing and the overall amplification depends on the total number of e-folds.} Note that $B_{0}$ is measured in Gauss and $M$ -- the scale of inflation -- in GeV. Setting $M\simeq10^{14}$~GeV, $1-\Omega_{0}\simeq10^{-2}$ and assuming a magnetic mode marginally inside the curvature radius, with $n\simeq1.01$ for example, we have\footnote{The size of the superadiabatically amplified magnetic seed is originally close to the present curvature scale of the universe. Based on the current WMAP data, the latter is no less than $10^{4}$~Mpc~\cite{Setal}. Nevertheless, the initial $B$-field is expected to break up and reconnect on much smaller lengths, when galaxy formation starts in earnest.}
\begin{equation}
B_{0}\sim 10^{-16} \hspace{5mm} \mathrm{G}\,,  \label{B02}
\end{equation}
In general, the greater the scale of inflation the stronger the magnetic amplification. On the other hand, the higher the density of the universe today, the weaker the final magnetic field. However, the $\Omega_{0}$-dependence is much weaker than the $M$-dependence, which means that even very marginally open universes can sustain astrophysically significant $B$-fields. For instance, setting $1-\Omega_{0}\sim10^{-10}$ and keeping $M\sim10^{14}$~GeV leads to $B_{0}\sim10^{-20}$~G near the curvature scale. This increases back to $B_{0}\sim10^{-16}$~G if we raise the scale of inflation to $10^{16}$~GeV.

Magnetic fields with the above strengths cannot affect nucleosynthesis, or leave a significant imprint in the cosmic microwave background, but can seed the galactic dynamo. In fact, one could even imagine a scenario where the superadiabatic amplification is strong enough to produce the $\mu$G-fields seen in galaxies without the need of the dynamo. For example, raising the scale of inflation to $10^{17}$~GeV and leaving $1-\Omega_{0}$ at $10^{-2}$ in Eq.~(\ref{B01}) gives $B_{0}\sim10^{-10}$~G today. Unless, we push inflation closer to the Planck scale, without too much gravitational wave production, this is probably the strongest magnetic field that can be obtained via our mechanism. Note that these values refer to the comoving $B$-field and do not include its subsequent amplification during the protogalactic collapse. The latter could add three or four orders of magnitude to the magnetic strength, especially when the more realistic scenario of an anisotropic collapse is adopted and the associated shearing effects are accounted for~\cite{DBL}. In that case, comoving magnetic fields of $10^{-10}$~G can in principle reach the $\mu$G level of their observed galactic counterparts without the need of a dynamo amplification.

What is most intriguing, however, is that magnetic fields around $10^{-15}$~G were recently reported in intergalactic voids by three independent groups~\cite{TGFBGC}. It is difficult to explain the presence of such fields by invoking late-time, post-recombination, mechanisms of magnetic generation. Thus, the plausible alternative is to look for a cosmological origin, notwithstanding the ambitious extrapolation that must be made from conditions in the very early universe to the final state of galaxies and clusters at late times. There are further obstacles, however, even when one goes beyond conventional physics. In the majority of the proposed mechanisms the $B$-field is generated and superadiabatically amplified during inflation. Afterwards, the adiabatic magnetic decay is restored and the field drops like $B\propto a^{-2}$ until today. As a result, many -- though not all -- of the related scenarios suffer from the so-called ``magnetic back-reaction'' problem. In other words, to achieve astrophysically relevant $B$-fields the inflationary amplification must be so strong that the magnetic energy density becomes comparable to that of the inflaton~\cite{DMR}. There are no back-reaction issues here. The energy density of the superadiabatically amplified field is always well below that of the dominant matter species. Despite this, the residual $B$-seed is strong enough to account for essentially all the large-scale magnetic fields observed in the universe today. This is achieved because the amplification is not confined to the inflationary era, but extends throughout the lifetime of the universe: from the beginning of inflation until today.

There are a number of attractive aspects in the geometrical mechanism of magnetic amplification outlined here (see also~\cite{TK} for details). Simplicity is the first of them. There is no need for complicated and exotic couplings between the various fields involved. Also, it is not necessary to break away from classical electromagnetic theory, to abandon standard cosmology, or to introduce any kind of new physics. It operates within conventional Maxwellian electromagnetism and Friedmannian cosmology. The only proviso is that our universe is marginally open today. Even very marginally open Friedmann cosmologies, with $1-\Omega_{0}\sim 10^{-10}$ or less, can sustain astrophysically relevant magnetic fields, with current magnitudes greater than $10^{-20}$~G.

One further consequence of the analysis we have presented suggests a further direction for future investigation. The universe contains a significant population of voids, regions with lower than average density present on the last scattering surface, which are significantly non-linear underdensity perturbations with $(\rho -\bar{\rho})/\bar{\rho}\lesssim -0.8$, where $\bar{\rho}$ is the mean density~\cite{BC}. These regions will evolve like 'small' open universes after the last scattering epoch and become increasingly spherical if they remain expanding in all directions and are not dominated by intrinsic gravitational wave anisotropies. Our analysis leads us to expect that the strongest residual magnetic fields may be found in these void regions. There, the decay of any primordial $B$-field would have been slowed down even further by the effects of the negative spatial curvature than in higher density domains, so long as the material content can support these fields.

In conclusion, we have shown that universes having $|1-\Omega_{0}|\lesssim10^{-2}$ have quite different cosmological magnetic field evolution, depending on whether $\Omega_{0}$ is greater than or less than unity. In a flat or marginally closed universe it is not possible for cosmological magnetic fields to survive from the inflationary era to the present, with strengths great enough to seed a dynamo, or explain the $\mu G$-order fields in high-redshift protogalaxies and the observed intergalactic field strengths. In contrast, marginally open universes can easily sustain primordial $B$-fields with strengths around $10^{-16}$~G. Observations of such magnetic fields may therefore be an indication that we live in an open universe with hyperbolic spatial geometry.

\end{document}